\documentstyle[amssymb,pre,aps,multicol,epsfig]{revtex}
\begin{document}

\begin{title}
{\bf Three-Dimensional Solitary Waves and Vortices in a  Discrete Nonlinear Schr{\"o}dinger Lattice}
\end{title}

\author{P.G.\ Kevrekidis$^{1}$, B.A.\ Malomed$^2$, D.J.\ Frantzeskakis$^{3}$
and R.\ Carretero-Gonz\'{a}lez$^4$}
\address{$^{1}$ Department of Mathematics and Statistics, University of
Massachusetts,
Amherst MA 01003-4515, USA \\
$^{2}$ Department of Interdisciplinary Studies, Faculty of Engineering,
Tel Aviv University, Tel Aviv 69978, Israel  \\
$^{3}$ Department of Physics, University of Athens,
Panepistimiopolis, Zografos, Athens 15784, Greece\\
$^4$ Nonlinear Dynamical Systems Group, Department of Mathematics,
San Diego State University, San Diego CA, 92182}

\date{To appear in Phys.\ Rev.\ Lett.\ (2004).}

\maketitle

\begin{abstract}
In a benchmark dynamical-lattice model in 
three dimensions, 
the discrete nonlinear Schr{\"{o}}dinger 
equation, we find discrete vortex solitons 
with various values of the topological charge $S$.
Stability regions for the vortices with $S=0,1,3$ are investigated. The $S=2$
vortex is unstable, spontaneously rearranging into a stable one with $S=3$.
In a two-component extension of the model, we find a novel class of stable
structures, consisting of vortices in the different components, perpendicularly oriented to each other.
Self-localized states of the
proposed types can be observed experimentally in Bose-Einstein condensates
trapped in optical lattices, and in photonic crystals built of microresonators.
\end{abstract}

\begin{title}
{\bf Three-Dimensional Solitary Waves and Vortices in a  Discrete
Nonlinear Schr{\"o}dinger Lattice}
\end{title}


\vspace{2mm}

\begin{multicols}{2}


{\it Introduction.} Intrinsic localized modes (ILMs), alias {\it
discrete breathers}, in nonlinear dynamical lattices have inspired
a vast array of theoretical and experimental studies. They have attracted
attention due to their inherent ability to concentrate and,
potentially, transport the vibrational energy in a coherent
fashion (see, e.g., \cite{review}). Settings in which these
objects appear range from arrays of nonlinear-optical waveguides
\cite{mora} and photonic crystals \cite{PhotCryst} to
Bose-Einstein condensates (BECs) in optical-lattice (OL)
potentials \cite{tromb}, and from various systems based on
nonlinear springs \cite{pesa} to Josephson-junction ladders
\cite{alex} and dynamical models of the DNA double strand
\cite{Peybi}.

A benchmark model, which generically emerges in the description of
dynamics in nonlinear lattices, is the discrete nonlinear
Schr{\"{o}}dinger (DNLS) equation \cite{DNLS}. It finds its most
straightforward physical realizations in two of the
above-mentioned settings, viz., arrays of optical waveguides
\cite{fibers,fiberExper}, and networks of BEC\ drops trapped in
OLs \cite{tromb}. While the former system may be, effectively,
$1$- and $2$-, but not $3$-dimensional, the latter was
experimentally created in three dimensions ($3$D) as well
\cite{greiner}, which suggests a direct physical implementation of
the $3$D DNLS model. Another physical realization of the $3 $D
DNLS equation may be provided by a crystal built of
microresonators trapping photons \cite{photons} or polaritons
\cite{polar}.

In spite of the physical relevance of the NLS equation in the $3$D
case, very few works attempted to find localized solutions in this
system, and those were actually done in the absence of OLs \cite{komineas}.
Only very recently, a possibility of the existence of stable $3$D solitons in continuum NLS equations including OL potentials has been demonstrated \cite{Salerno}. A problem of fundamental interest, both in its own right and in view of the possibility of the experimental realization (principally in the BEC-OL setting), concerns the search for $3$D solitons in the
DNLS model proper. Especially intriguing is a possibility to construct stable ILMs with 
{\em intrinsic vorticity} (topological charge), which would be an entirely new class of ILMs in $3$D.

In this work, we demonstrate that the discreteness indeed makes it possible
to stabilize, in the DNLS model with attraction, not only ordinary $3$D
ILMs, but also vortex solitons (they are strongly unstable in the continuum
limit;
notice, however, that $3$D vortex solitons can be stabilized in continuum
models with competing nonlinearities \cite{doughnut}). These include not
only fundamental discrete vortices, with the topological charge $S=1$, whose
stability in the $3$D case may be surprising by itself, but also
higher-order vortices, such as ones with $S=3$ (in the above-mentioned $3$D
continuum models with competing nonlinearities, stable higher-order vortices
have not yet been found). We also extend the considerations to
multi-component DNLS models, that allow for the existence and stability of
still more challenging configurations. In particular, we introduce a novel
type of a compound vortex, consisting of two vortices with the same value of
$S$ in the two components, whose orientations are {\em perpendicular} to
each other. Such solutions are stable too, in certain parametric intervals.

{\it The Model.} The DNLS equation on the cubic lattice with a coupling
constant $C$ is \cite{DNLS}
\begin{equation}
i\frac{d}{dt}\phi _{l,m,n}+C\Delta _{2}\phi _{l,m,n}+\left\vert \phi
_{l,m,n}\right\vert ^{2}\phi _{l,m,n}=0,  \label{NLS}
\end{equation}with $\Delta _{2}\phi _{l,m,n}\equiv \phi _{l+1,m,n}+\phi _{l,m+1,n}+\phi
_{l,m,n+1}+\phi _{l-1,m,n}+\phi _{l,m-1,n-1}+\phi _{l,m,n-1}-6\phi _{l,m,n}$.
We seek for ILM solutions $\phi _{l,m,n}=\exp (i\Lambda t)u_{l,m,n}$,
where $\Lambda $ is the frequency (chemical potential in the BEC context)
and the stationary eigenfunctions $u_{l,m,n}$ obey the equation
\begin{equation}
\Lambda u_{l,m,n}=C\Delta _{2}u_{l,m,n}+\left\vert u_{l,m,n}\right\vert
^{2}u_{l,m,n}.  \label{standing}
\end{equation}
Due to the invariance of Eq.\ (\ref{NLS}) against the scaling
transformation, $t\rightarrow t/U^{2}$, $C\rightarrow CU^{2}$, and
$u\rightarrow u/U$, with an arbitrary constant $U$, one can either
fix the coupling constant, as $C\equiv C_{0}$, and vary $\Lambda
$, with the objective to explore a full family of solutions of a
certain type, or, alternatively, fix $\Lambda \equiv \Lambda
_{0}$, and follow the variation of $C$. The actual control
parameter, that is invariant against the scaling, is
$C/\Lambda$.

Solutions to Eq.\ (\ref{standing}) (generally, complex ones) are obtained by
means of a Newton method. To test their stability, perturbed solutions are
used in the form \cite{joh}
\begin{equation}
\phi _{l,m,n}=e^{i\Lambda t}\left[ u_{l,m,n}+\epsilon \left(
a_{l,m,n}e^{\lambda t}+b_{l,m,n}e^{\lambda ^{\ast }t}\right) \right] ,
\label{pert}
\end{equation}
where $\epsilon $ is an infinitesimal amplitude of the
perturbation, and $\lambda $ is its (generally, complex)
eigenfrequency. The substitution of Eq.\ (\ref{pert}) into Eq.\
(\ref{NLS}) gives rise to linearized equations for the
perturbation eigenmodes,
\[
i\lambda \left(
\begin{array}{c}
a_{k} \\
b_{k}^{\ast }\end{array}\right) =\left(
\begin{array}{cc}
\partial F_{k}/\partial u_{j} &~ \partial F_{k}/\partial u_{j}^{\ast } \\[1.2ex]
-\partial F_{k}^{\ast }/\partial u_{j} &~ -\partial F_{k}^{\ast
}/\partial u_{j}^{\ast }\end{array}\right) \left(
\begin{array}{c}
a_{k} \\
b_{k}^{\ast }\end{array}\right) ,
\]
where $F_{k}\equiv
-C(u_{k+1}+u_{k-1}+u_{k+N}+u_{k-N}+u_{k+N^{2}}+u_{k-N^{2}}-6u_{k})+\Lambda
u_{k}-\left\vert u_{k}\right\vert ^{2}u_{k}$, and the string index,
$k$, is defined so that
$(l,m,n) \mapsto k\equiv l+(m-1)N+(n-1)N^2$.
Dirichlet boundary conditions were
imposed.

According to the scale invariance discussed above, we examine solutions
of Eq.\ (\ref{standing}) by fixing the frequency, $\Lambda =2$,
and varying the coupling $C$ 
(in the 
BEC context, this
means fixing the chemical potential, and varying the OL strength,  
as is experimentally feasible).
The solutions with different values of the topological charge $S$
(here, it ranges between $0$ and $3$) 
are generated by an
iterative scheme with an initial ansatz motivated by 
$3$D vortices in the continuum limit \cite{doughnut},
\begin{eqnarray}
u_{l,m,n}^{({\rm init})} &=&A[(l-l_{0})+i(m-m_{0})]^{S}\exp \left(
-|n-n_{0}|\right)  \\
&&\times {\rm sech}\left( \eta \sqrt{(l-l_{0})^{2}+(m-m_{0})^{2}}\right) ,
\end{eqnarray}
where $(l_{0},m_{0},n_{0})$ is the location of the vortex' center,
and $\eta $ is a scale parameter. The Newton algorithm was then
iterated until it converged to $1$ part in $10^{7}$. Our results
are typically shown for $9\times\! 9\times\! 9$ and $11\times\! 11\times\!
11$ site lattices, but larger ones were also investigated.

\begin{figure}[tbp]
\begin{center}
\begin{tabular}{lll}
~~~(a) ${\rm Re}(u_{l,m,n})=+0.5$ &  & 
~~~(b) ${\rm Re}(u_{l,m,n=5,6})$ \\[0.5ex] 
\epsfxsize=4.2cm \epsfysize=3.7cm \epsffile{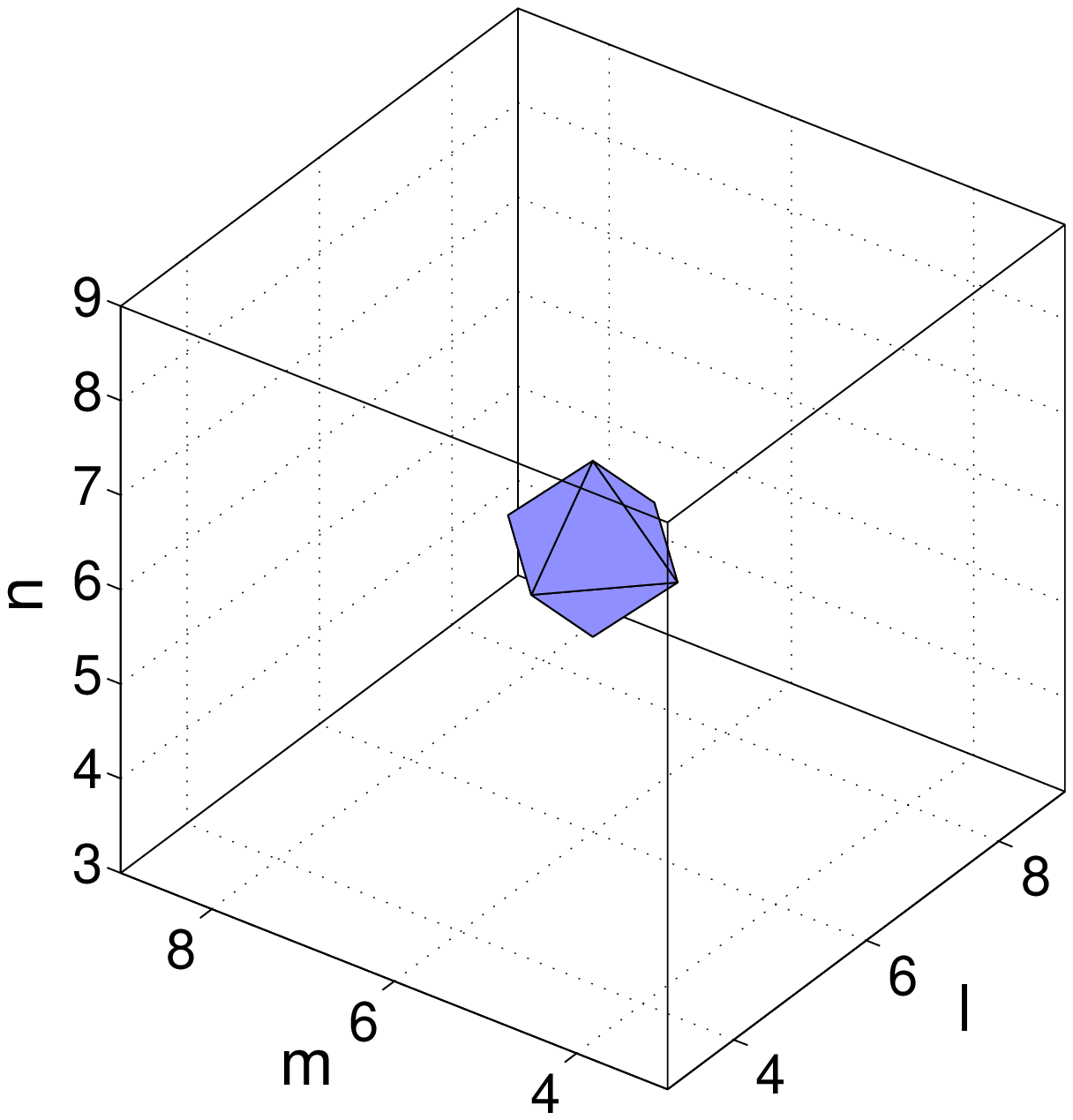} & & 
\epsfxsize=4.2cm \epsfysize=3.7cm \epsffile{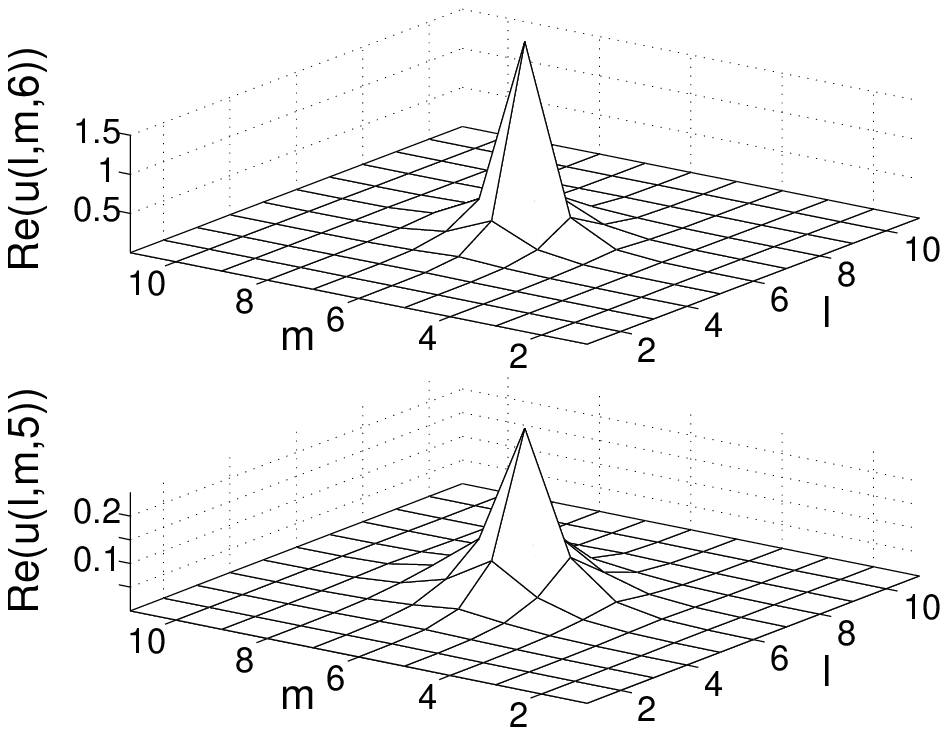}\\[-1.0ex]
\end{tabular}
\end{center}
\caption{The ILM with $S=0$ is shown for $C=1$. The left panel
shows the $3$D contour plot ${\rm Re}(u_{l,m,n})=0.5$.
The right panels show $2$D cross sections of the solution through
$n=6$ (top) and $n=5$ (bottom) for the $11\times\! 11\times\! 11$ DNLS
lattice. } \label{vfig1}
\end{figure}

{\it Results}. Basic results for the solutions with different
topological charges can be summarized as follows. Ordinary ILMs
with $S=0$ are stable below a critical value $C_{{\rm cr}}^{(0)} \approx 2$
of the coupling constant, as they are strongly unstable (against
collapse) in the continuum limit of $C\rightarrow \infty$.
An example of a stable ordinary soliton is
shown in Fig.\ \ref{vfig1}. As ILMs with $S=0$ have the largest
stability
interval, $C<C_{{\rm cr}}^{(0)}$, 
as compared to topologically charged 
ones (see below), they can only be destroyed if unstable, rather than being
transformed into ILMs of other types.

\begin{figure}[tbp]
\begin{center}
\begin{tabular}{lll}
~~~(a) ${\rm Re}(u_{l,m,n})=\pm 0.5$ &  & 
~~~(b) ${\rm Im}(u_{l,m,n})=\pm 0.5 $ \\[0.5ex]
\epsfxsize=4.2cm \epsffile{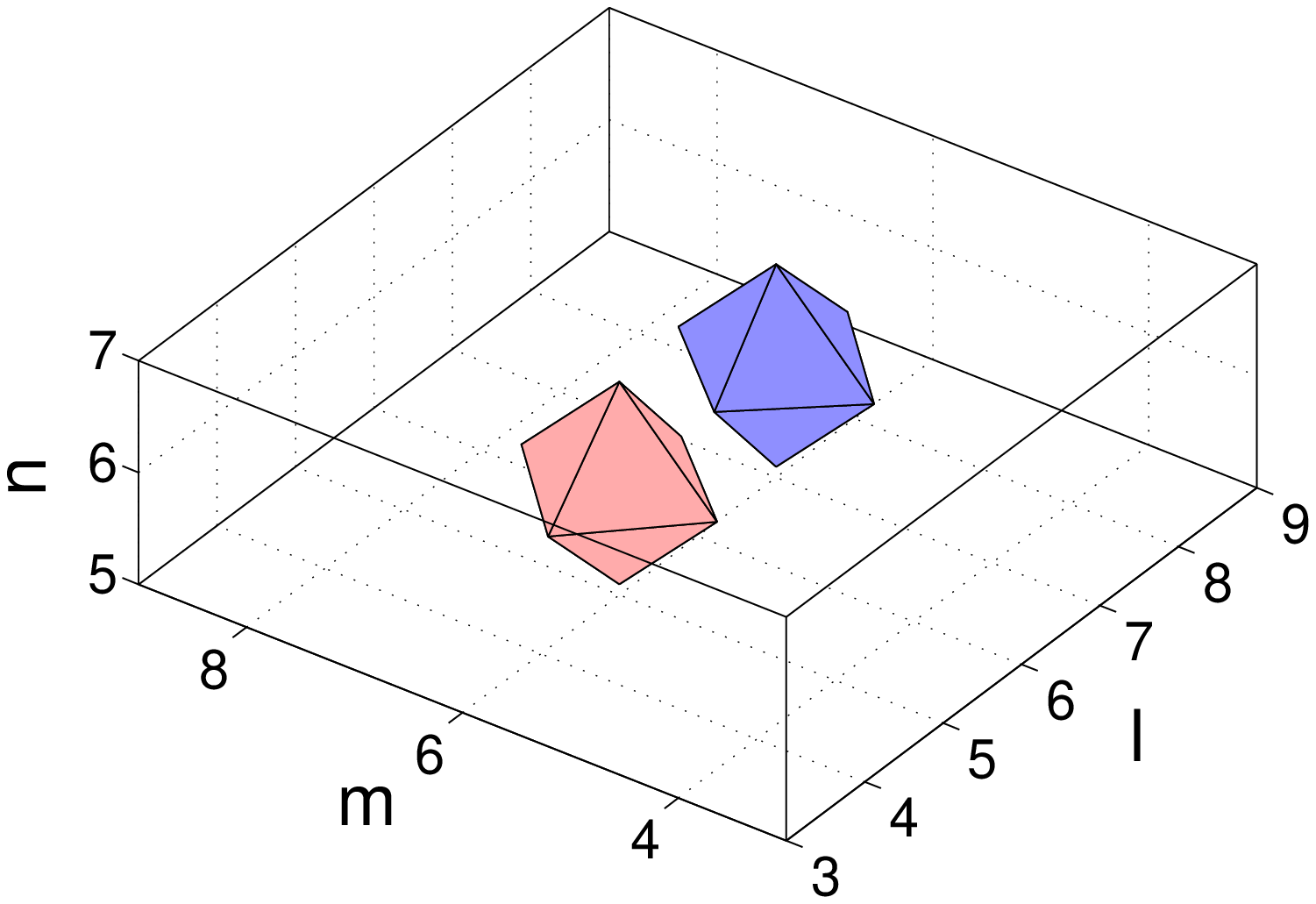} &  & 
\epsfxsize=4.2cm \epsffile{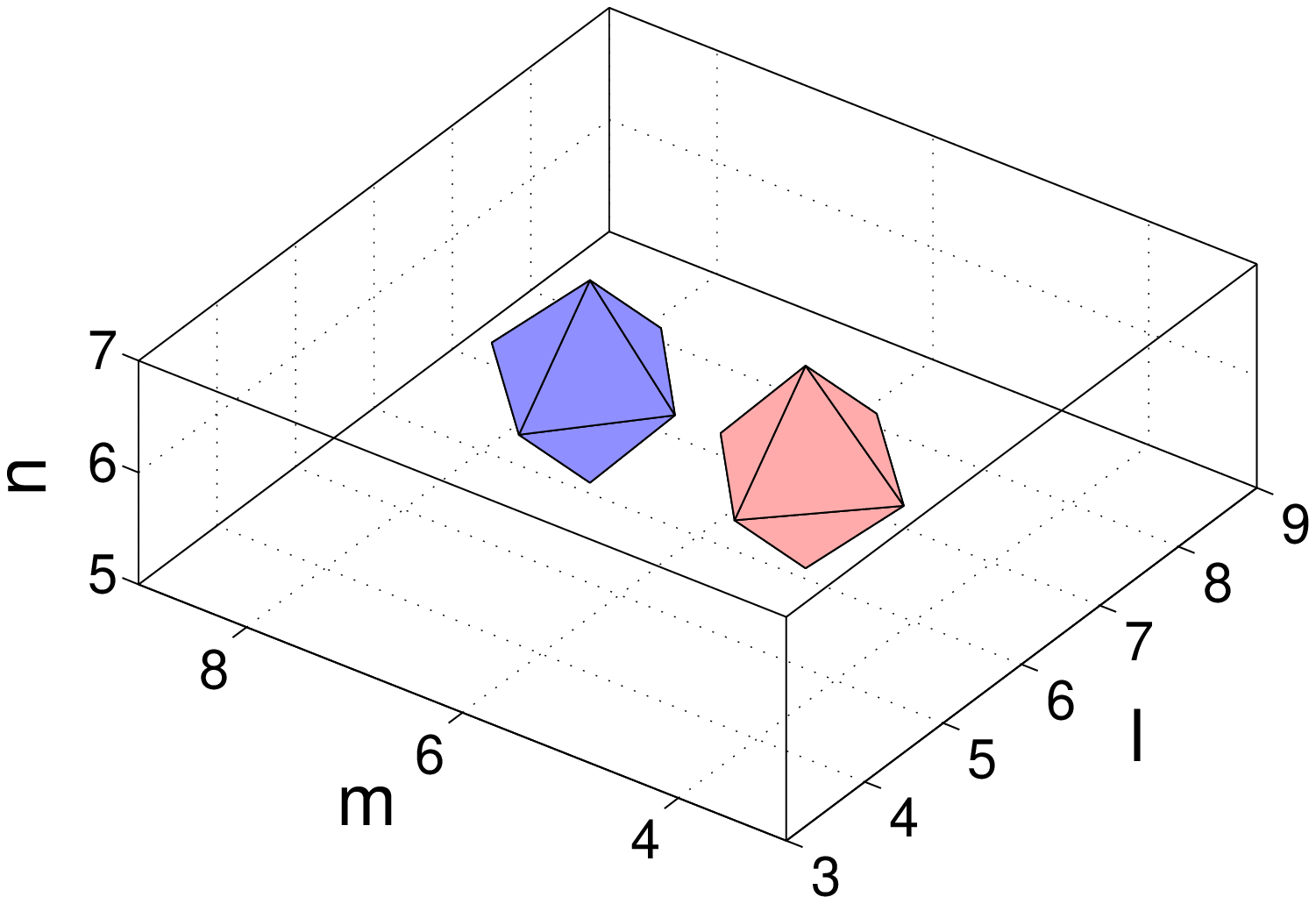} \\[-1.0ex]
~~~~~~(c) &  & ~~~~~~(d) \\[-1.0ex]
\epsfxsize=4.2cm \epsfysize=4.0cm \epsffile{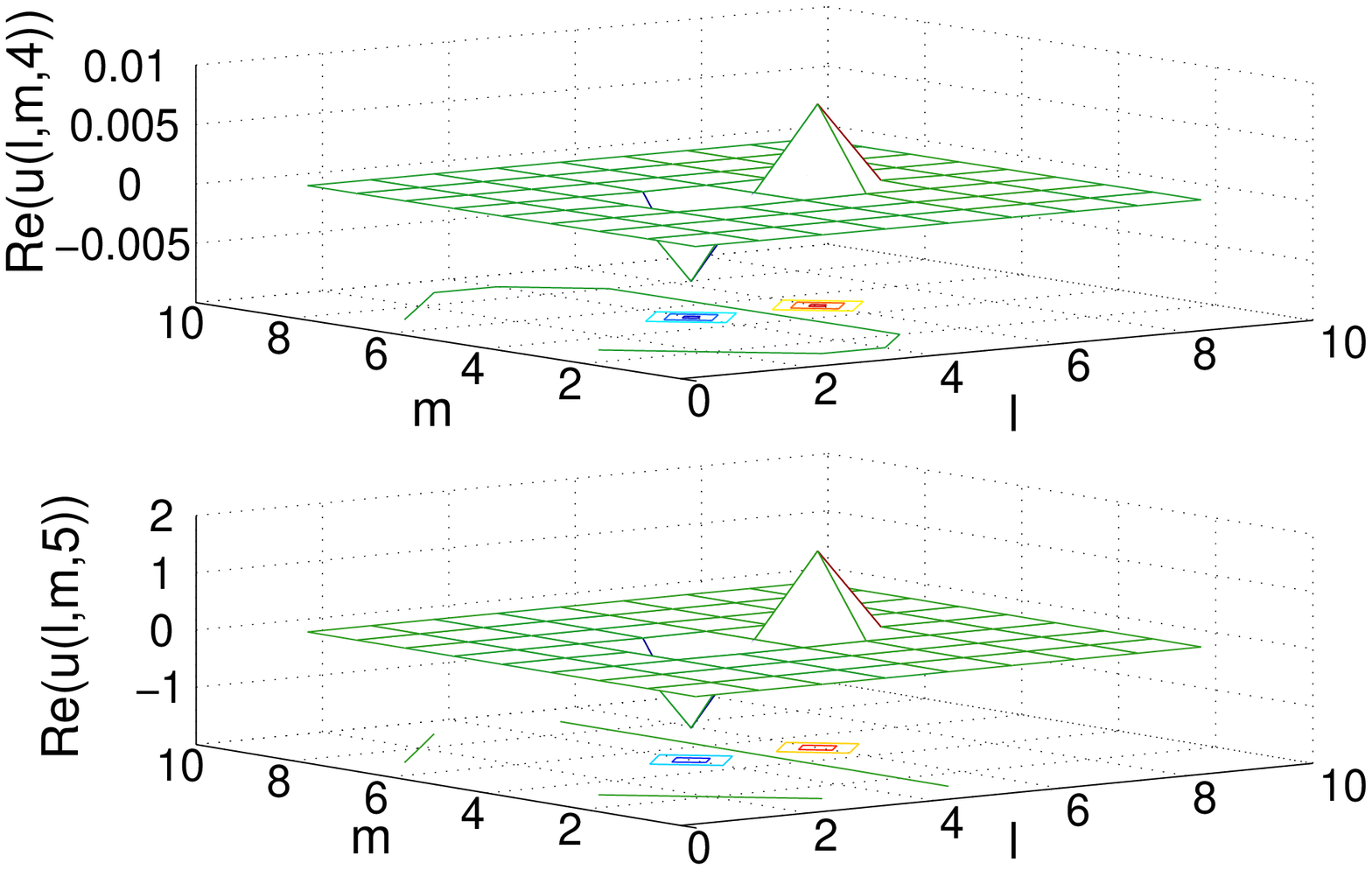} &  &
\epsfxsize=4.2cm \epsfysize=4.0cm \epsffile{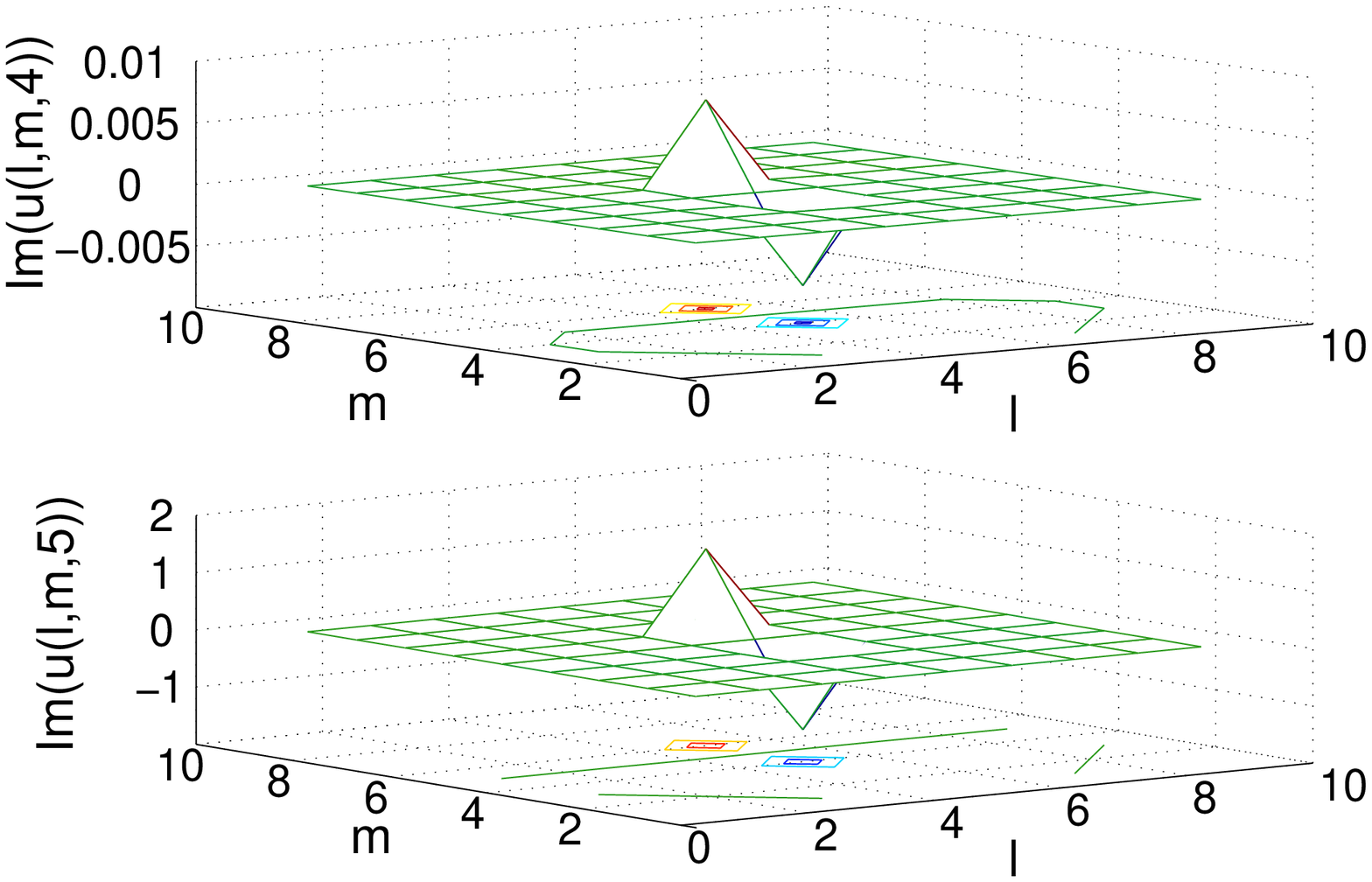} \\[-0.0ex]
~~~~~~(e) &  & ~~~~~~(f) \\
\epsfxsize=4.2cm \epsfysize=4.0cm \epsffile{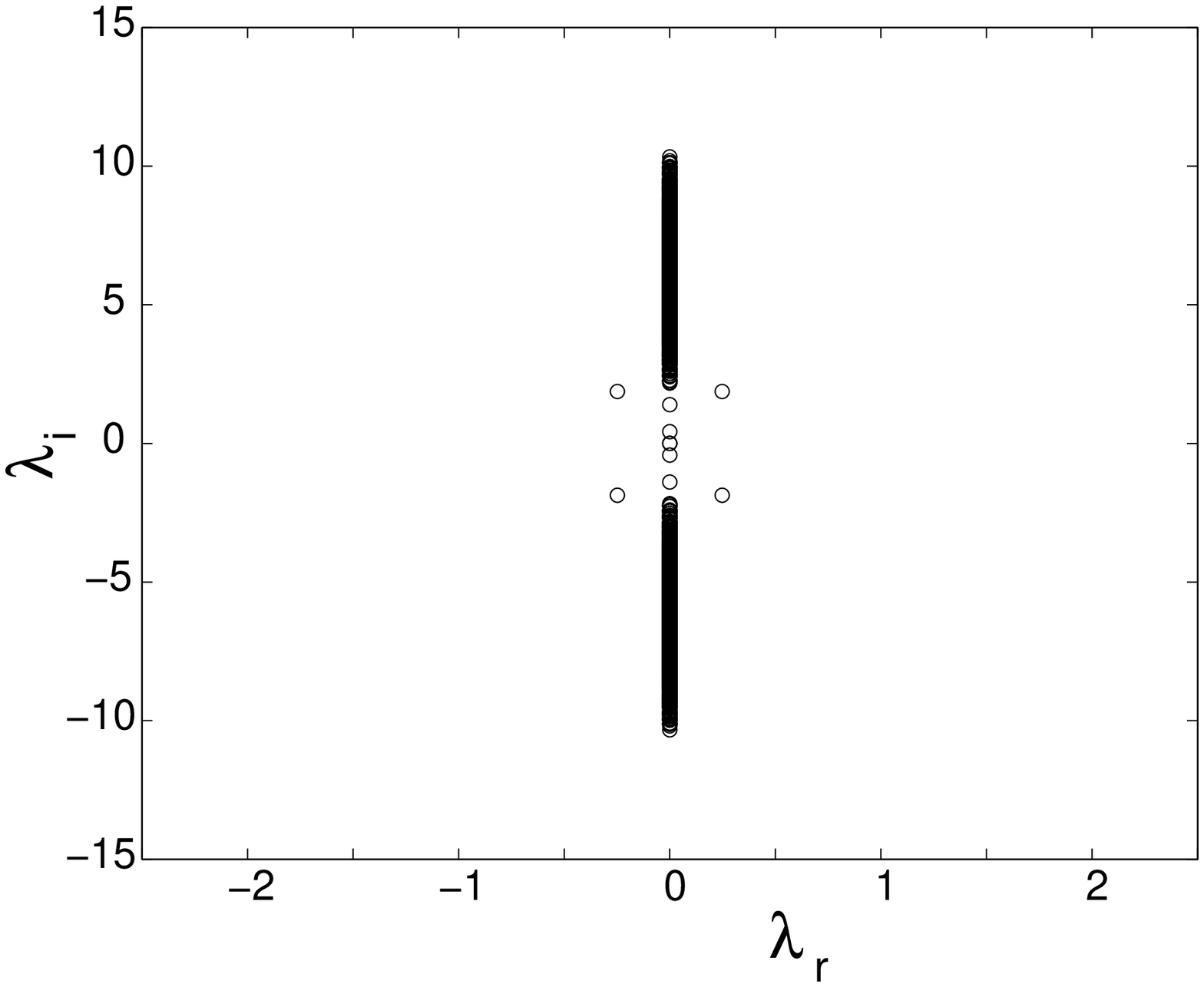} &  &
\epsfxsize=4.2cm \epsfysize=4.0cm \epsffile{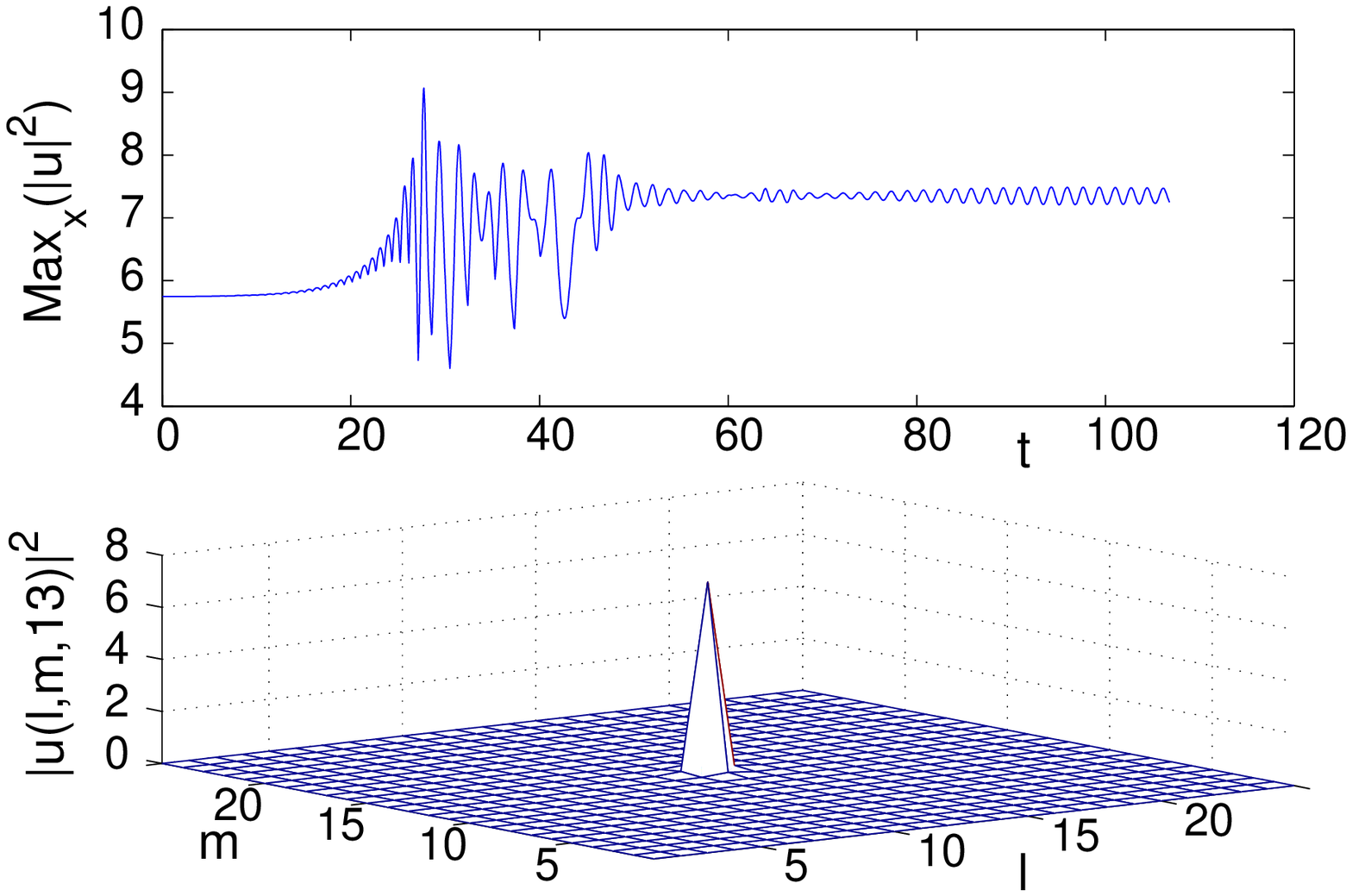} \\[-2.0ex]
\end{tabular}
\end{center}
\caption{The top panels show level contours at ${\rm
Re}(u_{l,m,n})=\pm 0.5$ (left) and ${\rm Im}(u_{l,m,n})=\pm 0.5$
(right) for the 3D vortex ILM with $S=1$. Red (color version) /
light-gray (black-and-white) and blue/dark-gray surfaces pertain
to the levels with $-0.5$ and $+0.5$ values, respectively. Cross
sections of the vortex are shown in four middle panels, (c) and
(d). The bottom row displays the development of instability of the
vortex for $C=0.7$, through the time evolution of its amplitude,
and a $2$D cut of the profile at $t=100$ [(f) top and bottom,
respectively]. The unstable vortex transforms itself into an
ordinary ILM\ with $S=0$. The left bottom panel (e) shows the
spectral plane $(\protect\lambda_{r},\protect\lambda_{i})$ of the
linear stability eigenvalues for the same unstable vortex.}
\label{vfig2}
\end{figure}

$3$D vortices with $S=1$ have also been found. They are stable
(see Fig.\ \ref{vfig2}) below a critical value $C_{{\rm
cr}}^{(1)} \approx 0.65$ (similarly to their 2D counterparts \cite{pgk4}).
At the instability threshold, a quartet of complex eigenvalues
emerges from collision of two imaginary eigenvalue pairs (for
details, see, e.g., Refs.\ \cite{pgk4,newpgk}). Numerically
simulated development of the instability is displayed in Fig.\
\ref{vfig2}, for a typical case with $C=0.7>C_{{\rm cr}}^{(1)}$.
The perturbations destroy the vortex structure and, as a result,
an ordinary ($S=0$) ILM emerges; obviously, the change of the
topological charge is possible in the lattice, in which the
angular momentum is not a dynamical invariant.

\begin{figure}[tbp]
\begin{center}
\begin{tabular}{lll}
~~~(a) ${\rm Re}(u_{l,m,n})=\pm 0.25$ &  & 
~~~(b) ${\rm Im}(u_{l,m,n})=\pm 0.25$ \\[0.5ex]
\epsfxsize=4.2cm \epsffile{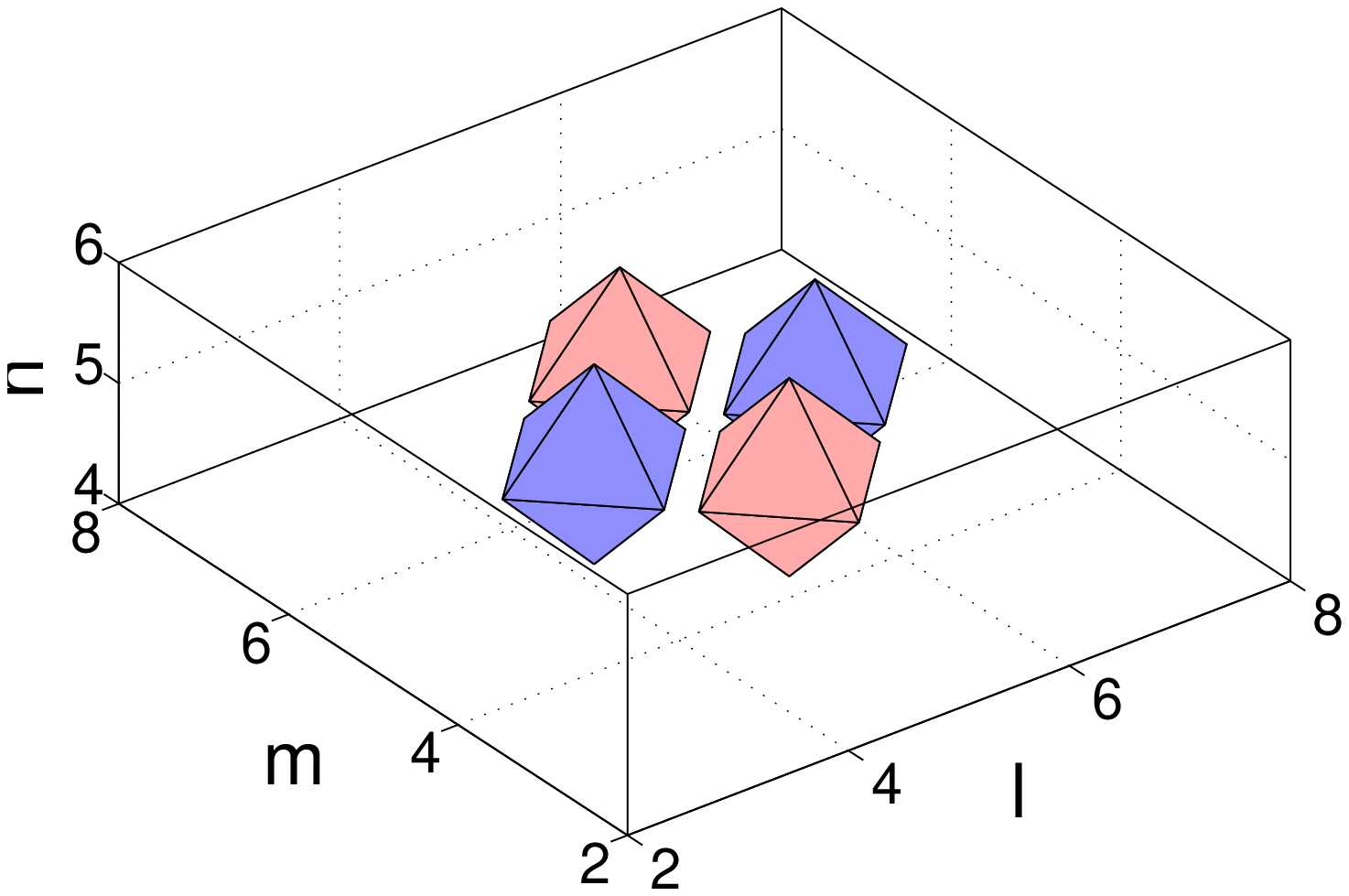} &  & 
\epsfxsize=4.2cm \epsffile{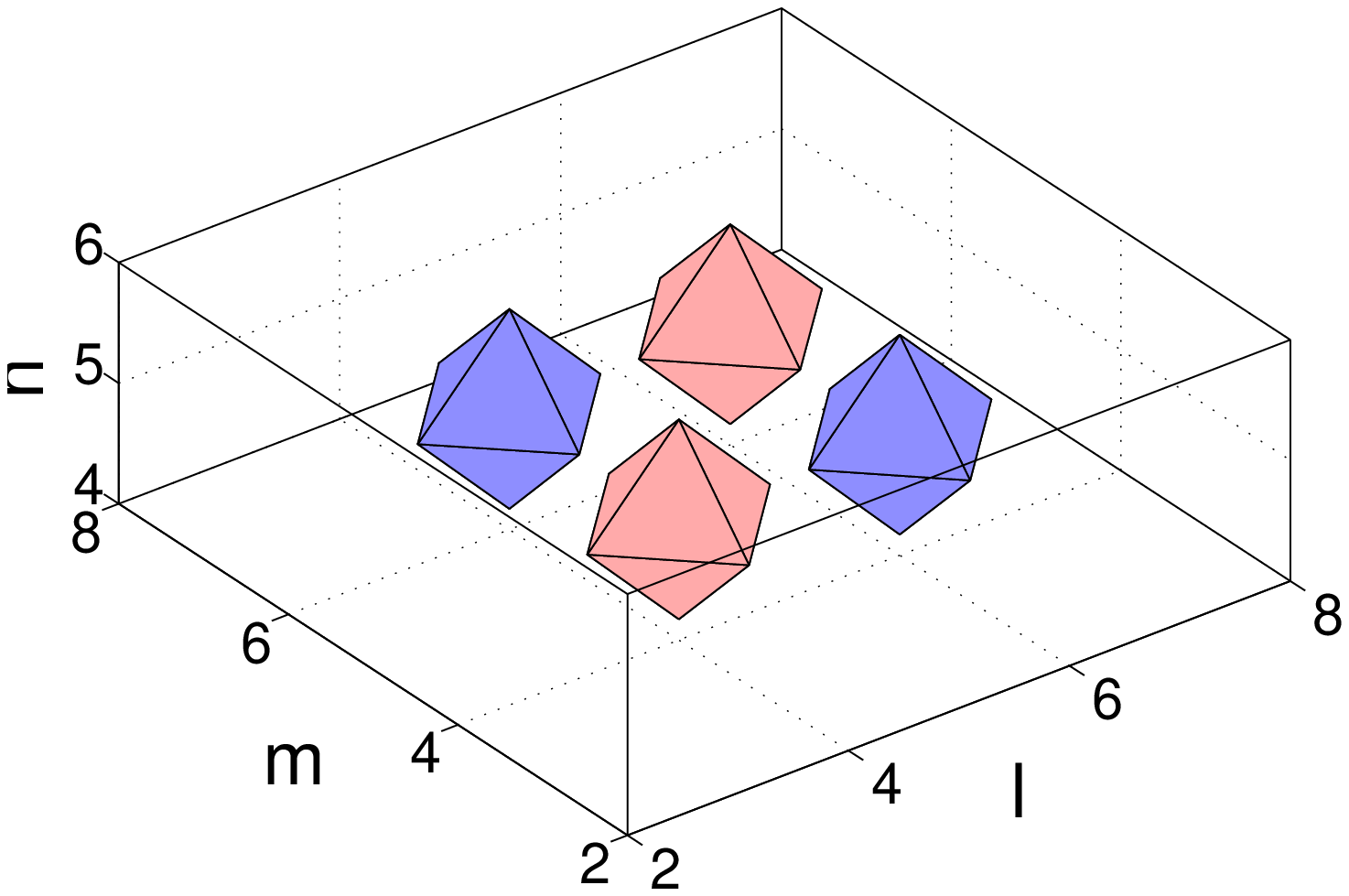} \\[-1.0ex]
~~~~~~(c) &  & ~~~~~~(d) \\
\epsfxsize=4.2cm \epsfysize=4.1cm \epsffile{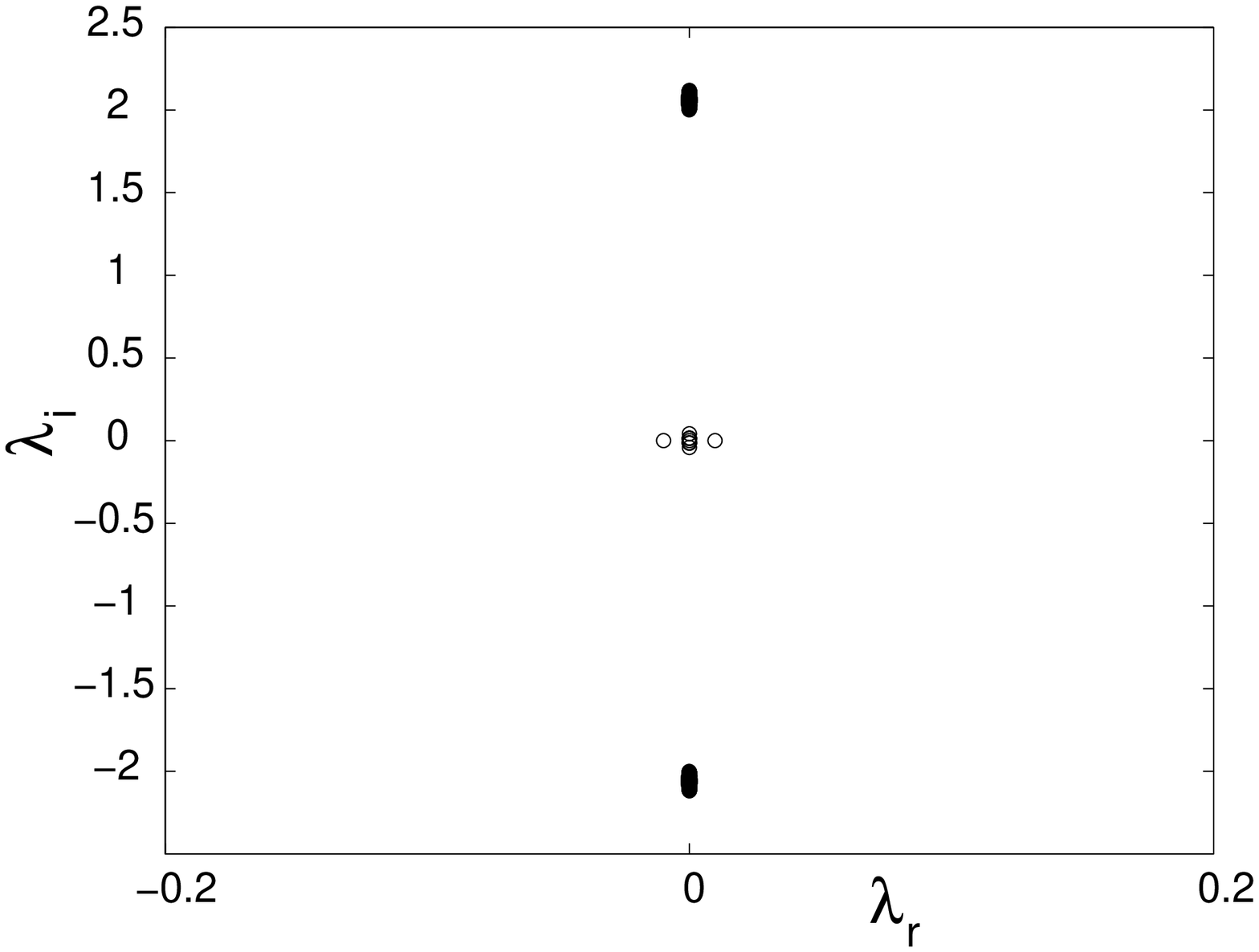} &  &
\epsfxsize=4.2cm \epsfysize=4.1cm \epsffile{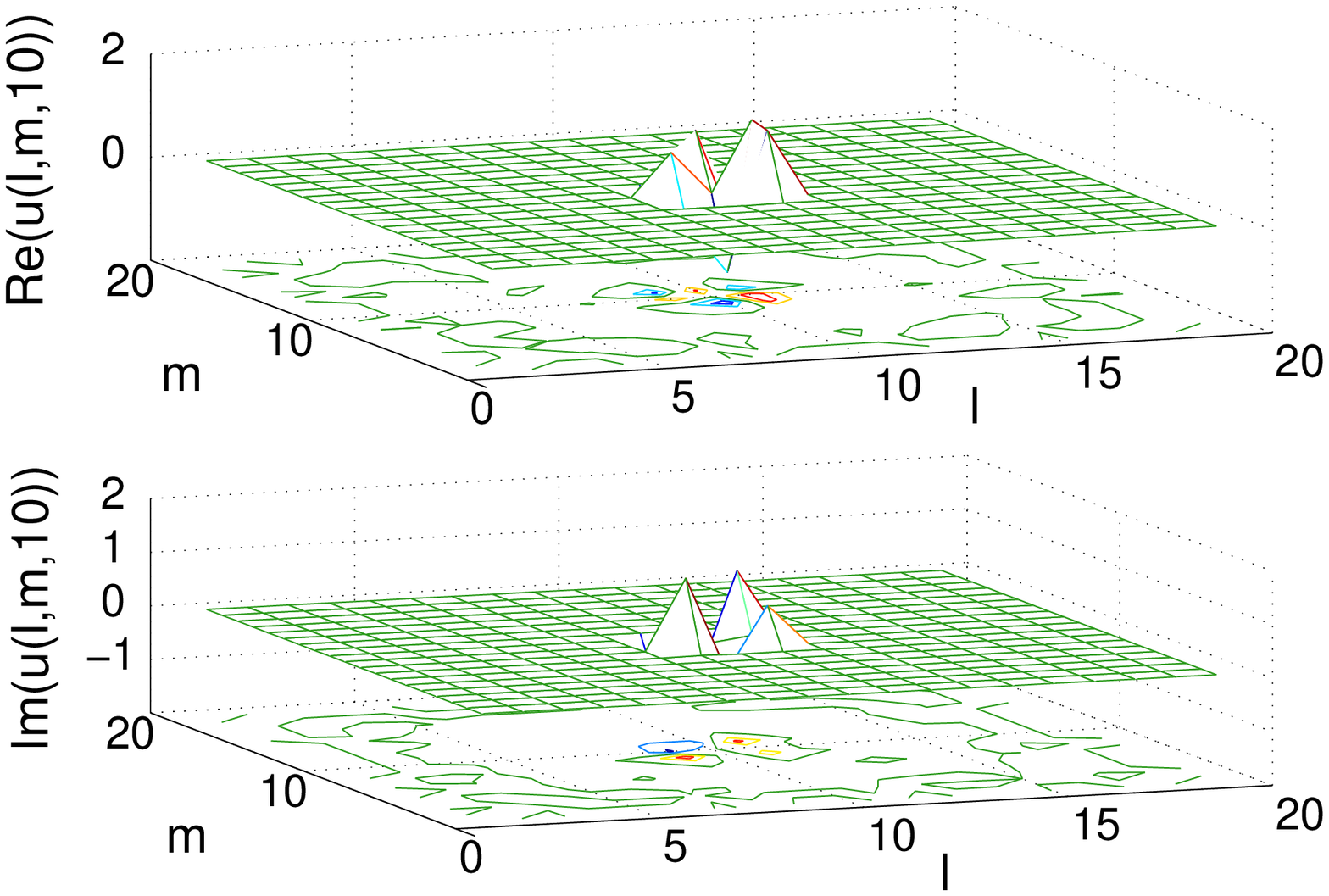} \\[-2.0ex]
\end{tabular}
\end{center}
\caption{The ILM vortex with $S=2$ for $C=0.01$. Top panels have the same
meaning as the top panels in Fig.\ \protect\ref{vfig2}. The bottom left
panel (c) displays the linear stability eigenvalues, while (d) shows the
result of long evolution of this unstable vortex. The eventual state, shown
through its $2$D cross-sections at $t=1000$, is a vortex with $S=3$
(see Fig.\ \ref{vfig4}), which
is {\em stable} for this value of $C$.}
\label{vfig3}
\end{figure}

\begin{figure}[tbp]
\begin{center}
\begin{tabular}{lll}
~~~(a) ${\rm Re}(u_{l,m,n})=\pm 0.25$ &  & 
~~~(b) ${\rm Im}(u_{l,m,n})=\pm 0.25$ \\[0.5ex]
\epsfxsize=4.2cm \epsffile{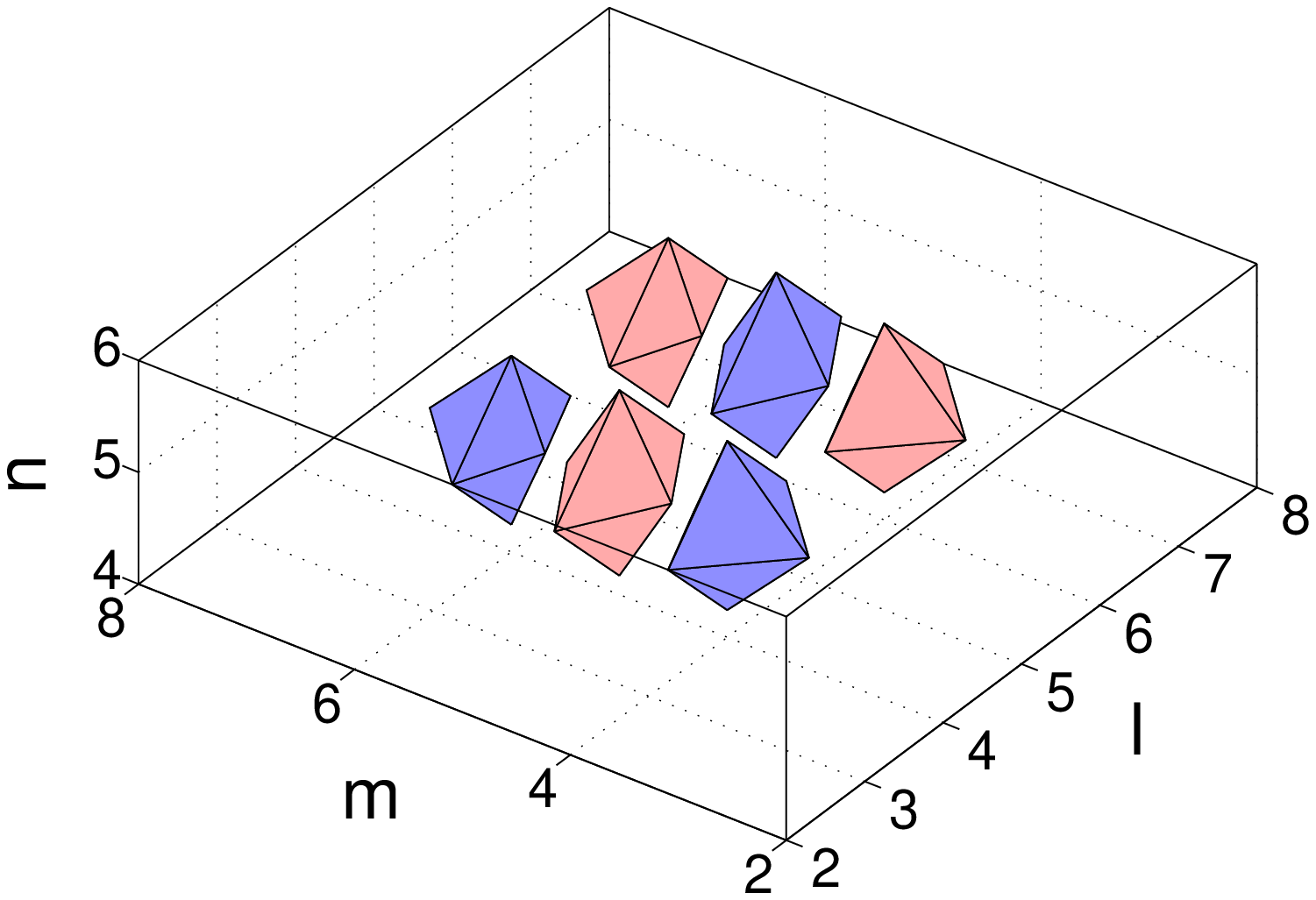} &  & 
\epsfxsize=4.2cm \epsffile{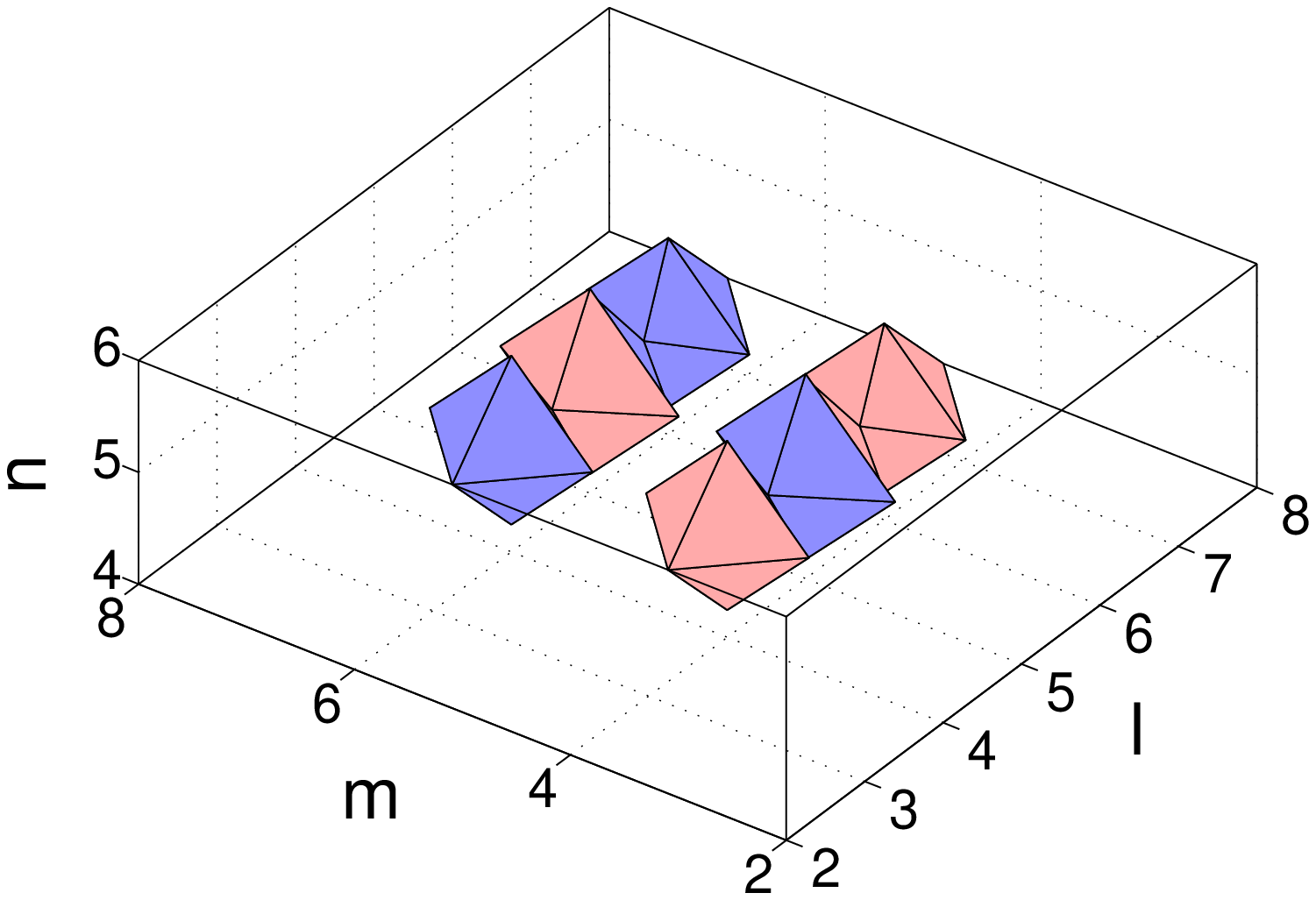}\\[-2.0ex]
\end{tabular}
\end{center}
\caption{Stable stationary vortex ILMs with $S=3$ for $C=0.01$. Panels
have same meaning as top panels in Fig.\ \protect\ref{vfig2}.}
\label{vfig4}
\end{figure}

An example of a vortex with topological charge $S=2$ is shown in
Fig.\ \ref{vfig3} for $C=0.01$. Similar to its $2$D counterpart
\cite{pgk4}, this complex solution is unstable through a real
eigenvalue pair at all values of $C$ (notice, however, that purely
real {\em stable} solutions to the 2D DNLS equation, that may be
identified as {\it quasi-vortices} similar to the solitons with
$S=2$, have been very recently found \cite{newpgk}; counterparts
of such solutions exist in the $3$D case as well). What is more
interesting, however, is that this unstable ILM with $S=2$
reshapes itself {\em not} into the one with $S=0$ or $S=1$, but
rather towards a stable vortex with $S=3$, as seen in Fig.\
\ref{vfig3}. The stabilization of the 3D vortex ILM through
spontaneous {\em increase} of $S$ is a striking result (again,
feasible only in dynamical lattices). In Fig.\ \ref{vfig4} we show
the stable $S=3$ discrete vortex for the same case, $C=0.01$
(stable higher-order vortex ILMs were very recently found in the
$2$D\ DNLS model too \cite{newpgk}). The instability of the vortex
with $S=2$ vs.\ the stability of the vortex with $S=3$ may be
understood, in loose terms 
(in the $2$D case as well) in terms of
the lattice-induced Peierls-Nabarro (PN) potential acting on the
soliton. Indeed, it is the PN potential which may stabilize a
soliton which is otherwise strongly unstable. It is seen from
Figs.\ \ref{vfig3} and \ref{vfig4} that the PN potential induced
by the cubic lattice is, obviously, a much stronger factor for the
soliton with $S=3 $ than with $S=2$, due to the symmetry
difference between the former one and the lattice (i.e., the
``skeleton'' of the vortex lies at angles of 
$2 \pi /3$ as opposed to the 
$\pi/2$ of the underlying cubic lattice).

\begin{figure}[tbp]
\begin{center}
\begin{tabular}{lll}
~~~(a) ${\rm Re}(\phi_{l,m,n})=\pm 0.25$ &  & 
~~~(b) ${\rm Im}(\phi_{l,m,n})=\pm 0.25$ \\[0.5ex]
\epsfxsize=4.2cm \epsfysize=3.7cm \epsffile{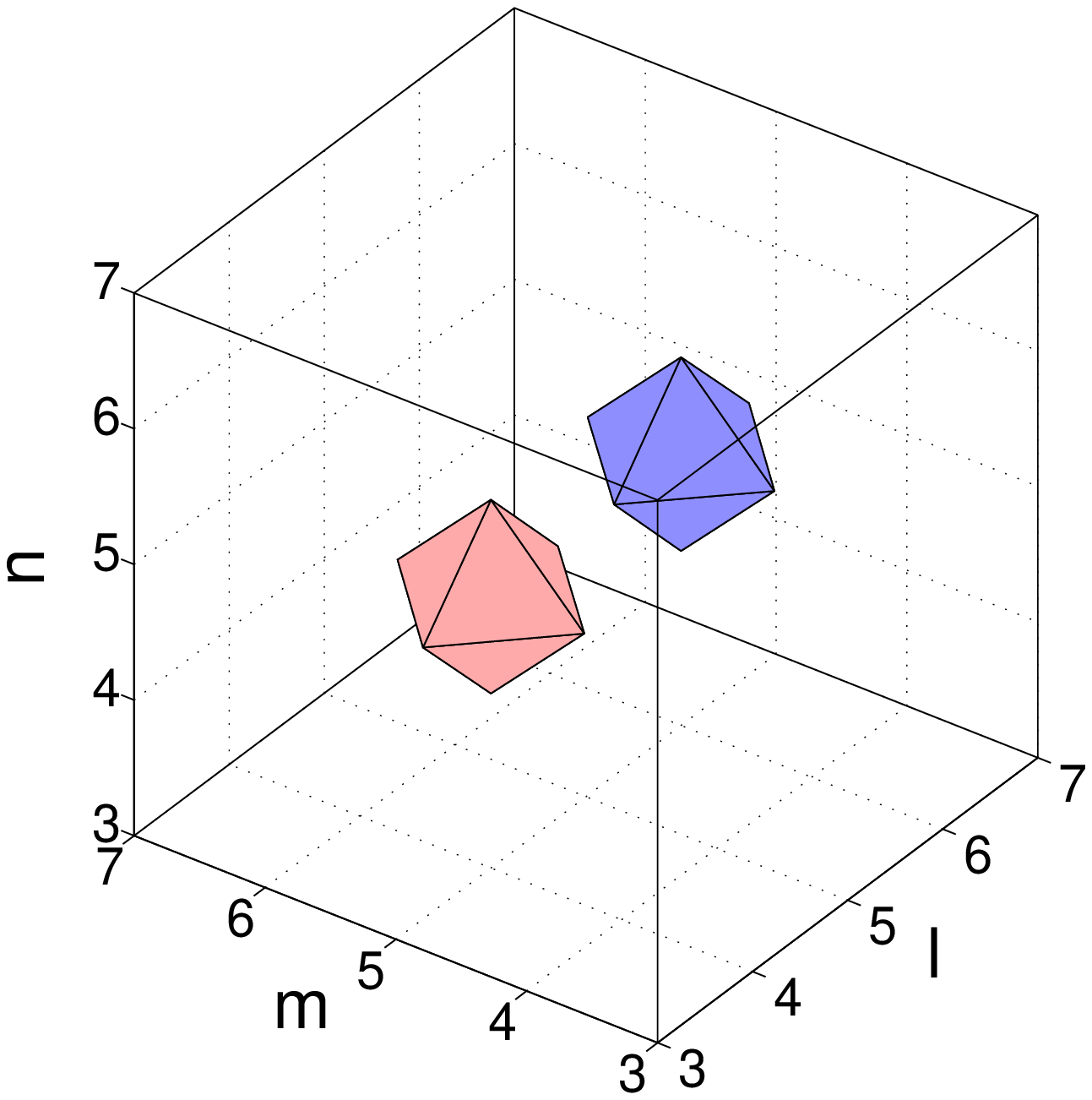} & &
\epsfxsize=4.2cm \epsfysize=3.7cm \epsffile{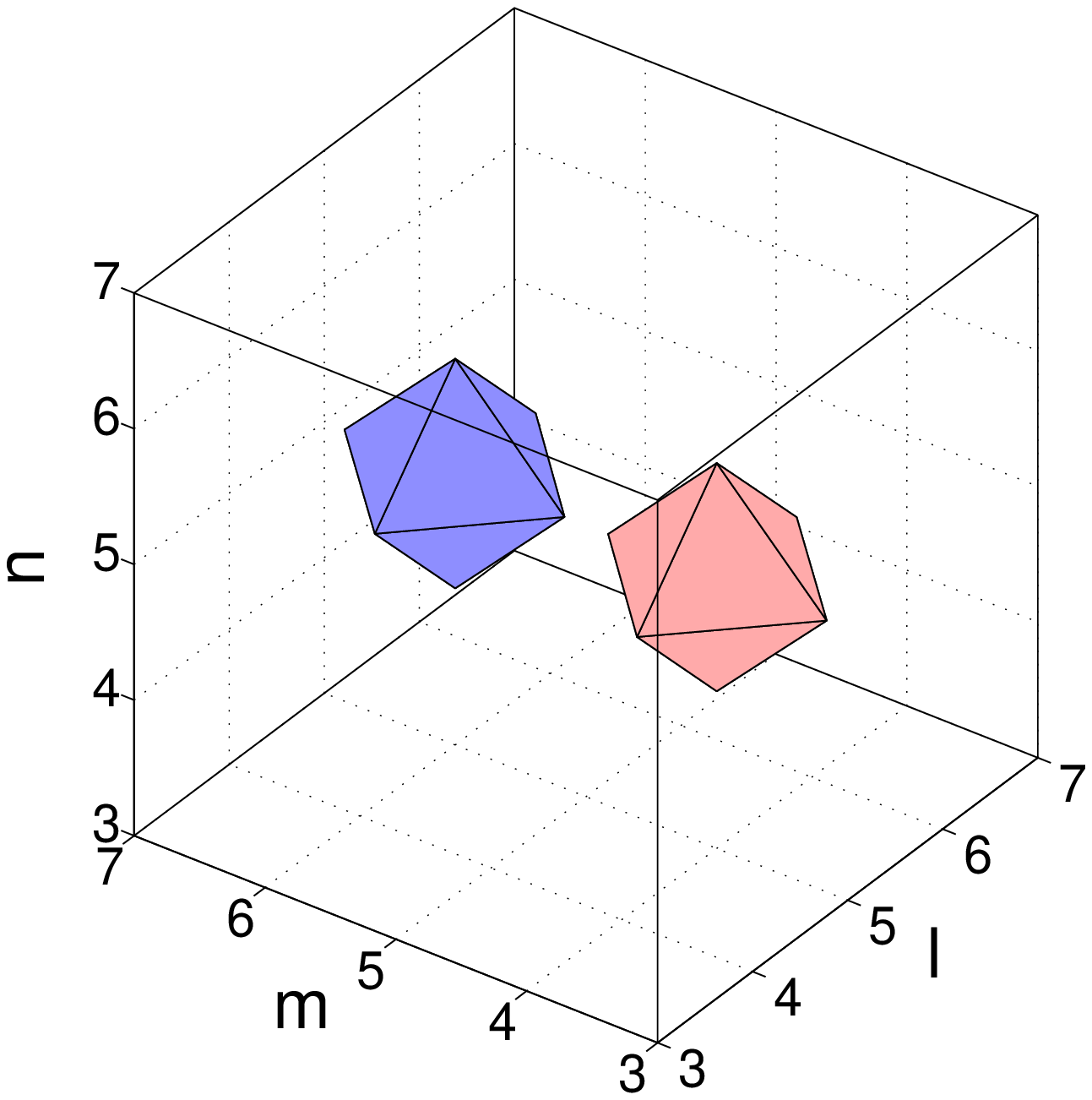} \\[0.5ex]
~~~(c) ${\rm Re}(\psi_{l,m,n})=\pm 0.25$ &  & 
~~~(d) ${\rm Im}(\psi_{l,m,n})=\pm 0.25$ \\[0.5ex]
\epsfxsize=4.2cm \epsfysize=3.7cm \epsffile{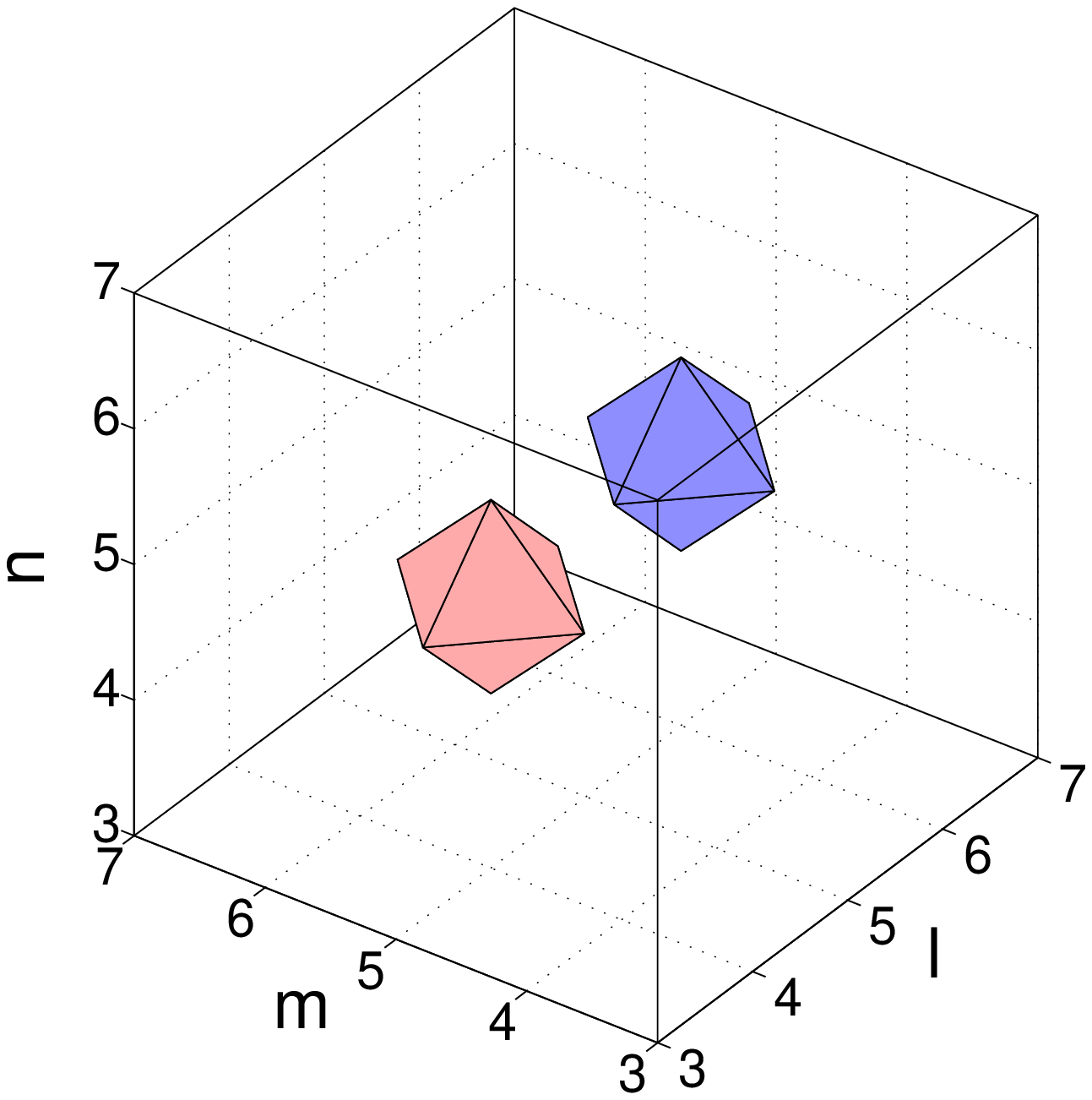} & &
\epsfxsize=4.2cm \epsfysize=3.7cm \epsffile{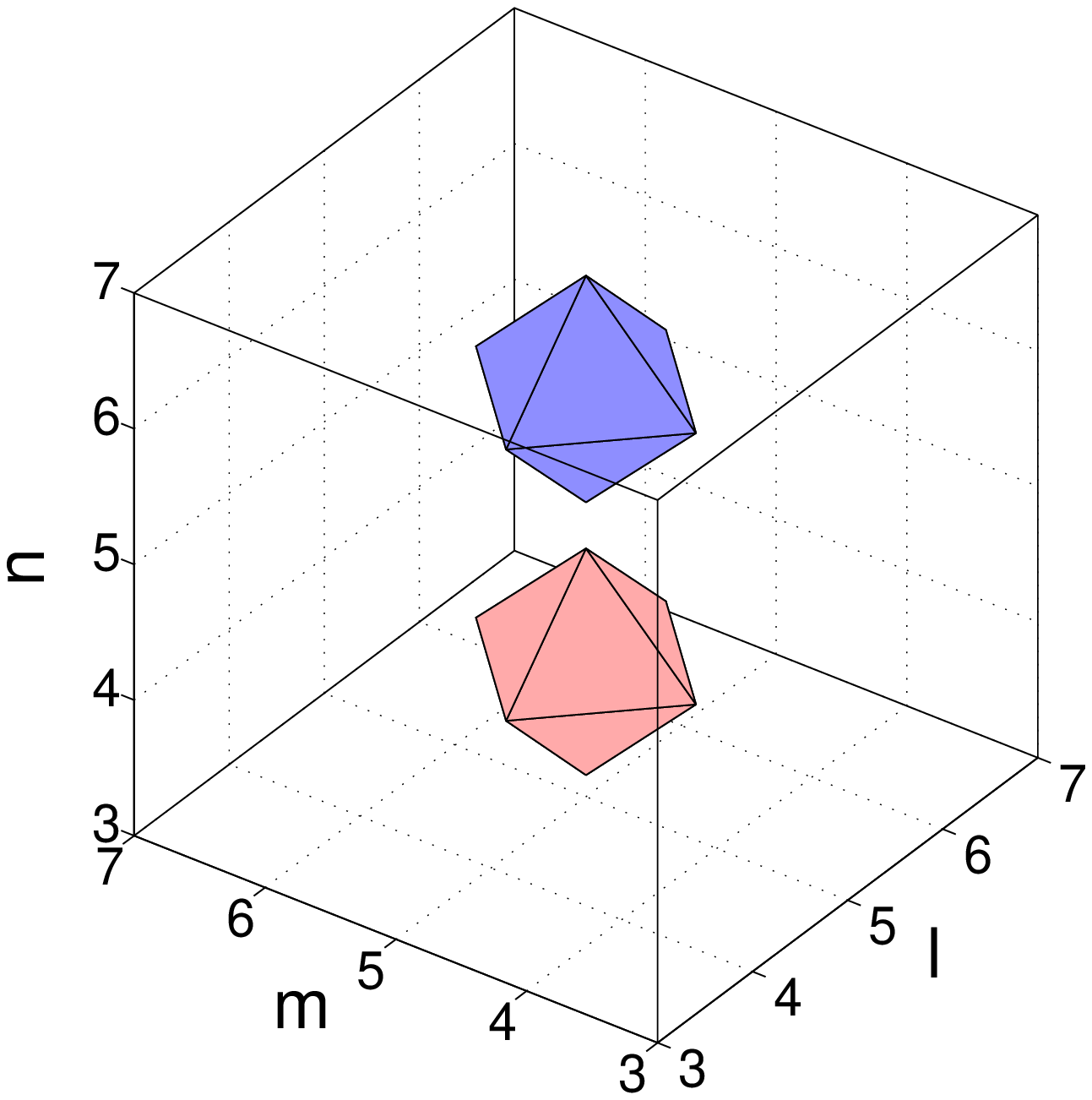}\\[-2ex]
\end{tabular}
\end{center}
\caption{A complex of two orthogonal vortices with $S=1$ in the
two-component system is shown for $C=0.01$. The top and bottom panels
correspond to the two components, and they have the same meaning as the top
panels in Fig.\ \protect\ref{vfig2}.}
\label{vfig5}
\end{figure}

Another striking feature, unique to the $3$D case, is a possibility of the
existence of vortex complexes in a multi-component system, with the vortices
in different (up to three) components {\em orthogonal} to each other. We
consider, in particular, two coupled DNLS equations,
\begin{equation}
\begin{array}{l}
\left[ i\frac{d}{dt}+C\Delta_2+ \left( |\phi_{l,m,n}|^{2}+\beta |\psi
_{l,m,n}|^{2}\right) \right] \phi _{l,m,n} =0 \\[1.0ex]
\left[ i\frac{d}{dt}+C\Delta_2+ \left( |\psi _{l,m,n}|^{2}+\beta
|\phi _{l,m,n}|^{2}\right) \right] \psi _{l,m,n} =0,
\label{phipsi}\end{array}\end{equation}
which describe an array of BEC droplets composed of a mixture of
two different species \cite{myatt}. In the case of the
model based on the photon or polariton field 
trapped in the lattice of microresonators, $\phi $ and $\psi $
refer to two different polarizations or distinct cavity modes. In
Eqs.\ (\ref{phipsi}), $\beta $ is the relative, intra-species interaction strength.

We examine a complex of two orthogonal vortices, in which the one
in the first component is directed perpendicular to the $\left(
l,m\right) $ plane, while in the second component, the vortex is
orthogonal to the $\left( l,n\right) $ plane. An example of such a
{\em stable} complex is shown, for $\beta =0.5$, in Fig.\
\ref{vfig5}. We have found that the orthogonal complexes may be
stable for $\beta <1$ (for sufficiently weak coupling $C$: the 
solutions of Fig.\ \ref{vfig5} are stable for $C<C_{cr}^{\perp} \approx 0.025$), 
and are unstable for $\beta >1$. This can be
qualitatively understood in terms of the Hamiltonian of the
attractive interaction between the two components, each having the
characteristic ``doughnut" \cite{doughnut} vortex-soliton shape
(in the continuum limit). Indeed, one can\ roughly estimate the
interaction energy (negative) through the volume $V$ of the overlap between
two cylinders of a radius $\rho $ (which represent long inner
holes of the doughnuts) intersecting at an angle $\theta $,
$V=(20/3)\rho ^{3}/\sin \theta $ (the divergence at $\theta
\rightarrow 0$ is limited by a finite length of the holes). As it
follows from here, the interaction energy has a maximum at $\theta
=\pi /2$,
which corresponds, by itself, to an unstable equilibrium
state of two orthogonal vortices. Actually, the equilibrium is
transformed into a stable one by the pinning to the PN potential,
provided that the interaction is not too strong, i.e., $\beta $ is
not too large.

{\it Conclusions.} The above results are a first step towards an
understanding of topologically nontrivial ILMs
in $3$D dynamical lattices. Besides generating stable ILMs without ($S=0$)
or with ($S=1$ and $3$) topological charge, the $3$D lattice gives rise to a
variety of novel dynamical effects, such as the reshaping of the $S=2$
unstable vortex into a stable one with $S=3$. Furthermore, the $3$D
dynamical lattice sustains quite unusual but stable states, such as the
two-component complex, with the individual vortices in the components
orthogonal to each other. Studying more complex configurations in such a
rich setup, 
and examining interactions between ILMs, are
challenging problems for future work.

The $3$D vortices predicted in this work can be created in an
self-attractive BEC trapped in an optical lattice (OL). Relevant
physical parameters are essentially the same as those at which
this medium is experimentally available
\cite{tromb,greiner,myatt}, and for which various $1$D and $2$D
localized structures were predicted \cite{tromb,Salerno}, i.e.,
$\gtrsim 10^4 $ atoms trapped in an OL with the period $\sim 1$
$\mu$m and the size $\gtrsim 10 \times\! 10 \times\! 10$. The
soliton-formation time is $\sim 1$ ms, and the experiment can be
easily run on the time scale of $\gtrsim 1$ s. In principle, the
same vortex solitons can be also created in a cubic lattice
composed of microresonators trapping photons or polaritons.

We gratefully acknowledge the support of NSF-DMS-0204585, NSF-CAREER, and
the Eppley Foundation for Research (PGK); 
the SDSU Foundation (RCG); and
the Israel Science Foundation grant No.\ 8006/03 (BAM).



\end{multicols}

\end{document}